\documentclass[aps,prb,superscriptaddress,reprint]{revtex4-2}

\usepackage[utf8]{inputenc}
\usepackage{xcolor}
\usepackage{pst-node}
\usepackage{graphicx}
\usepackage{amsmath, amssymb, amsthm}
\usepackage{mathtools}
\usepackage{hyperref}
\usepackage[shortlabels]{enumitem}
\usepackage{dsfont}
\usepackage{amsthm}
\usepackage{amssymb}
\usepackage{amstext}
\usepackage{amsfonts}
\usepackage{nicefrac}
\usepackage{extarrows} 
\usepackage{graphicx}
\usepackage{relsize}
\usepackage{soul}

\newcommand{\ket}[1]{\vert #1 \rangle}
\newcommand{\id}{\mathbb{I}}
\DeclareMathOperator{\Tr}{Tr}

\begin{document}

\title{Two dimensional quantum lattice models via mode optimized \\hybrid CPU-GPU density matrix renormalization group method} 

\author{Andor Menczer}
\affiliation{%
Strongly Correlated Systems Lend\"ulet Research Group,
Wigner Research Centre for Physics, H-1525, Budapest, Hungary
}%

\author{Kornél Kap\'as}
\affiliation{%
Strongly Correlated Systems Lend\"ulet Research Group,
Wigner Research Centre for Physics, H-1525, Budapest, Hungary
}%
\affiliation{%
Department of Theoretical Physics, Institute of Physics, Budapest University of Technology and Economics, M\H uegyetem rkp. 3, H-1111 Budapest, Hungary
}

\author{Mikl\'os Antal Werner}
\affiliation{%
Strongly Correlated Systems Lend\"ulet Research Group,
Wigner Research Centre for Physics, H-1525, Budapest, Hungary
}%

\author{\"Ors Legeza}
\email{legeza.ors@wigner.hu }
\affiliation{%
Strongly Correlated Systems Lend\"ulet Research Group,
Wigner Research Centre for Physics, H-1525, Budapest, Hungary
}%
\affiliation{
Institute for Advanced Study,Technical University of Munich, Germany, Lichtenbergstrasse 2a, 85748 Garching, Germany
}

\date{\today}

\begin{abstract}
\noindent \textbf{Abstract.} 
We present a hybrid numerical approach to simulate quantum many body problems on two spatial dimensional quantum lattice models via the non-Abelian ab initio version of the density matrix renormalization group method on state-of-the-art high performance computing infrastructures. We demonstrate for the two dimensional spinless fermion model and for the Hubbard model on torus geometry
that altogether several orders of magnitude in computational time can be saved by performing calculations on an optimized basis and by utilizing
hybrid CPU-multiGPU parallelization. At least an order of magnitude reduction in computational complexity results from mode optimization, while a further order of reduction in wall time is achieved by massive parallelization. Our results are measured directly in FLOP and seconds. A detailed scaling analysis of the obtained performance as a function of matrix ranks and as a function of system size up to $12\times 12$ lattice topology is discussed. Our CPU-multiGPU model also tremendously accelerates the calculation of the one- and two-particle reduced density matrices, which can be used to construct various order parameters and trace quantum phase transitions with high fidelity.
\end{abstract}

\maketitle

\section{Introduction}
\label{sec:intro}

Simulation of strongly correlated quantum many body problems in two- and higher spatial dimensions poses great challenge and is still part of active research in modern condensed matter physics~\cite{Zheng-2017,Arovas-2022,Qin-2022}. In spite of great efforts to generalize the density matrix renormalization group (DMRG) method~\cite{white-1992a,Schollwock-2005,Noack-2005, Hallberg-2006, Verstraete-2023}, that is the most accurate algorithm for one-dimensional systems, there is still no universal tensor network state (TNS) solution that could be applied in higher spatial dimensions with an efficiency approaching the DMRG method for one-dimensional systems. 

In fact, the DMRG method has serious limitations in higher spatial dimensions as even models with local interactions become longer ranged when mapped to the one-dimensional DMRG topology~\cite{Liang-1994, Noack-1994,White-2003}. Other approaches like the projected entangled pair states (PEPS) and its variants~\cite{Verstraete-2004a, Verstraete-2008,Orus-2014} or the tree tensor network state (TTNS) algorithms ~\cite{Murg-2010a,Tagliacozzo-2009,Gunst-2018,Gunst-2019,Nakatani-2013} are based on tensor networks that are already higher dimensional~\cite{Orus-2014,Orus-2019}, but, on the other hand, they have much higher computational demands due to significant increase in scaling exponents with increasing accuracy. Other alternatives, that are not based on TNS wavefunctions, also have limitations. Quantum Monte-Carlo (QMC) and its extensions are hindered by the so-called negative sign problem ~\cite{Himeda-2002, Chang-2010, Assaad-2013, Kolorenvc-2011}, that arises generally for fermionic and frustrated models, and can only be avoided for special model parameters~\cite{Assaad-2013,Li-2019}. Other approaches like DMFT, DMET, etc. are based on a self consistent simplification of the many-body Hamiltonian and are therefore non-variational.~\cite{Georges-1996, Kotliar-2006, Knizia-2013, Zheng-2017} Nevertheless, besides their limitations, all the above mentioned methods have also their benefits and are commonly used -- sometimes complementing each other -- as state-of-the-art numerical methods for two and higher dimensional many body quantum systems.~\cite{Zheng-2017} 

Entanglement scaling law -- the so-called area law -- for two and higher dimensional systems also indicates serious limitations for MPS based methods~\cite{Eisert-2010, Stoudenmire-2012}. Most of the DMRG-based analysis, however, have been performed for lattice models with local interactions and represented by localized basis sets, i.e., lattice models~\cite{Calabrese-2004,Eisert-2010}. We also note that using periodic boundary conditions for such lattice models, which is usually beneficial due to the suppressed finite size effects, may results in even higher entanglement~\cite{Schollwock-2011}. To circumvent these difficulties, a commonly used approach in two dimensions is to consider systems with a rather narrow stripe or cylinder topology, where boundary conditions are periodic for the shorter side while open for the long side~\cite{Zheng-2017, Jiang-2020a,Huang-2022}. This approach is sometimes supplemented by a transformation to momentum space in the shorter direction to exploit also the translational invariance~\cite{Zheng-2017,Motruk-2016, Ehlers-2017}. In these calculations, however, the length of the shorter side of the system is usually rather limited ($L_\perp \lesssim 7-8$) that may results in stronger, and spatially inhomogeneous finite size effects.

The curse of high entanglement can be removed or at least reduced by transforming to a more general basis via non-local, one-particle unitary transformations. Using such non-local basis can alter scaling properties and bring in new ingredients especially for approximate, i.e., truncated solutions. An optimal unitary transformation can result in drastically decreased entanglement and, consequently, a large reduction of tensor ranks, also called the bond dimension~\cite{Krumnow-2016,Krumnow-2021}. In contrast, the non-locality of the transformation, can result in a Hamiltonian with long-ranged interactions instead of the original lattice model with short ranged couplings only. The combined effect of these two opposing mechanisms determines the total computational cost of the calculation, reduction of which, as much as possible, is mandatory for efficient simulations on classical computers.
\begin{figure}
    \centering
    \includegraphics[width=0.5\textwidth]
    {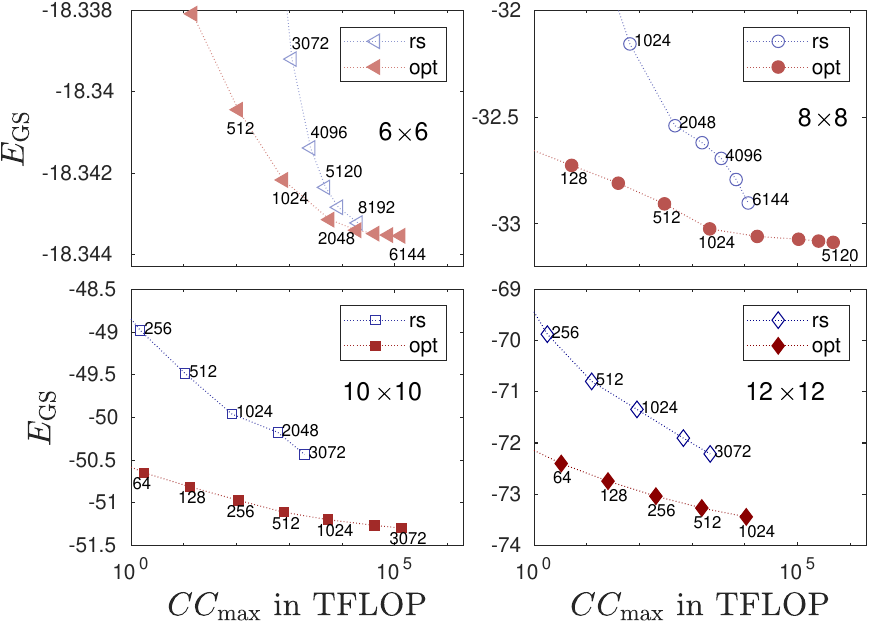}
\caption{Ground state energy as a function of the maximum computational complexity, $CC_{\rm max}$,
    for the half-filled, $n \times n$, $n=6,8,10,12$ two-dimensional spinless fermion model with $t=1, t^\prime= 0.4$ and $V=0.8$ on a torus geometry using real space (rs) and optimized (opt) basis. Numbers next to some selected data points label DMRG bond dimension.
    }
    \label{fig:E_CC_rs_opt}
\end{figure}

This work is devoted to demonstrate the efficiency of our approach to target low energy eigenstates of two- and higher spatial dimensional quantum many body systems by a generalized variant of the DMRG method. First the conventional real space basis is relaxed and a more general basis is employed and optimized via fermionic mode transformation~\cite{Krumnow-2016}. As a result, the computational complexity of the problem
(number of floating point operations), $CC$, for a given accuracy threshold can be reduced by an order of magnitude or even more. This is demonstrated in Fig.~\ref{fig:E_CC_rs_opt} for the two dimensional spinless fermion model at half filling on a torus geometry up to lattice sizes of $12\times 12$ with nearest and next-nearest neighbor hopping, $t=1$ and $t^\prime=0.4$, respectively, and nearest neighbor Coulomb interaction $V=0.8$. This point in the $t^\prime-V$ phase space was selected as a numerically very challenging one due to the vicinity of a quantum phase transition, as will be shown in Sec.~\ref{sec:spinless}. 
The tremendous reduction, which also manifests itself in drastic saving in wall time, is based on the minimization of the artificial entanglement and correlation introduced by mapping the physical system to a given tensor network topology~\cite{Krumnow-2021}. 

\begin{figure}
    \centering
    \includegraphics[width=0.48\textwidth]{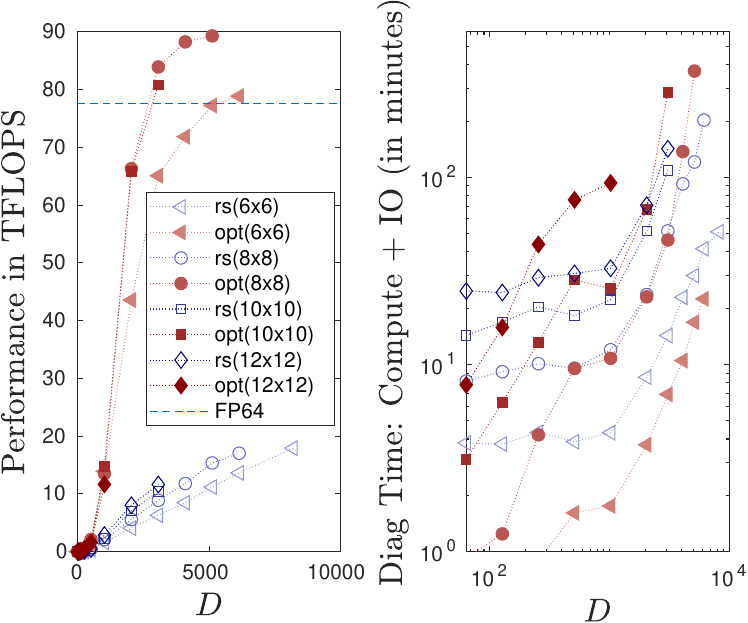}
    \caption{
    Maximum performance in TFLOPS (left) 
    and total diagonalization time including device to host (D2H) IO of nine DMRG sweeps in minutes (right)
    measured for the eight GPU accelerated diagonalization of the effective Hamiltonian as a function of the bond dimension using the real space and the optimized basis for model parameters given in Fig.~\ref{fig:E_CC_rs_opt}.
    The estimated FP64 theoretical upper bound for eight NVIDIA A100-PCIE-40GB GPU devices is shown by the dashed line. Note that achieved performance above the FP64 theoretical upper bound is due to utilization of the highly specialized NVIDIA tensor core units (TCUs). 
    }
    \label{fig:Perf_rs_opt}
\end{figure}

Next we show that another order of magnitude reduction in wall time can be achieved via massive parallelization utilizing state-of-the-art hardware and software solutions. Although the locality of the Hamiltonian is lost via mode optimization, leading to significant increase in the so called matrix product operator (MPO) bond dimension, our benchmark calculations using a hybrid CPU-multiGPU DMRG~\cite{Menczer-2023a,Menczer-2023b}
demonstrates that the underlying computational power 
of high performance architecture (HPC) 
can be utilized more efficiently for the optimized basis (see left panel of Fig.~\ref{fig:Perf_rs_opt}). Therefore, for a given DMRG bond dimension, the total wall time of the real and optimized basis can be brought much closer for a broad range of the bond dimension values, $D$, as is shown in the right panel of Fig.~\ref{fig:Perf_rs_opt}. 
Recalling from Fig.~\ref{fig:E_CC_rs_opt} that by using the optimized basis the same accuracy can be achieved with much smaller $D$ values, the overall reduction in wall time can be well over two orders of magnitude for a given accuracy (see, for example, $D=512$ data sets for the $8\times8$ lattice model). Our method, therefore, has the potential to open a new direction to study models in two- and higher spatial dimensions. 

The reduction in block entropy and the efficiency of mode optimization have already been analyzed in terms of the DMRG bond dimension for the spinless fermion model~\cite{Krumnow-2021}, but a detailed scaling analysis in terms of the underlying computational complexity ($CC$), i.e., the competition between the MPS bond and MPO dimensions, remained unexplored, although computational complexity is the purest measure that determines the efficiency of a numerical approach at the end. In addition, the suitability for massive parallelization can also bring in new ingredients when efficiency is discussed. Therefore, here we aim to fill this gap by measuring $CC$ directly in FLOP, wall time after massive parallelization in seconds and effect of $SU(2)$ spin symmetry on these quantities. 

The setup of the paper is as follows.  Sec.~\ref{sec:theo} is devoted for a brief overview of model systems and mode optimization. In Secs.~\ref{sec:spinless} and \ref{sec:hubbard} we present scaling analysis and numerical benchmark results obtained by large scale mode optimized hybrid CPU-multiGPU DMRG calculations for the two dimensional spinless fermion and for the Hubbard models respectively. Our work closes with main conclusions and future perspectives.

\section{Theory}
\label{sec:theo}
\subsection{Model Hamiltonians}

Our numerical framework to simulate properties of two dimensional quantum lattice models relies on 
the DMRG method~\cite{White-1992b}, that is a special variant of TNS algorithms ~\cite{Schollwock-2011,Noack-2005,Chan-2008,Szalay-2015a,Orus-2014,Baiardi-2020}, and on
a very general form of the Hamiltonian operator, 
\begin{equation}
\mathcal{H} = \sum_{ij\alpha\beta} T_{ij}^{\alpha\beta} 
                c^\dagger_{i\alpha}c_{j\beta} +
                \sum_{ijkl\alpha\beta\gamma\delta} 
      {V_{ijkl}^{\alpha\beta\gamma\delta}
      c^\dagger_{i\alpha}c^\dagger_{j\beta}c_{k\gamma}c_{l\delta}},
\label{eq:ham}
\end{equation}
implemented in our code ~\cite{dmrg-budapest},  
that can treat any form of non-local interactions related to two-particle scattering processes. 
The operators $c^\dagger_{i\alpha}$ or $c_{i\alpha}$ denote fermionic creation and annihilation operators \footnote{Operators $c^\dagger_{i\alpha}$ and $c_{i\alpha}$ can denote also spin ladder operators in our DMRG code but fermionic mode optimization can only be employed for fermionic systems.}, where indices $i$ and $\alpha$ specify mode and (generalized) spin information of the single particle state, respectively. 
The general form \eqref{eq:ham} of the Hamiltonian allows for simulation of strongly correlated quantum many body systems in various fields of disciplines, like condensed matter physics, nuclear structure theory or quantum chemistry even in the relativistic domain
~\cite{White-1992b,Xiang-1996,White-1999,Schollwock-2005,Noack-2005,Chan-2008,Schollwock-2011,Knecht-2014,Dukelsky-2004,Orus-2014,Legeza-2015,Szalay-2015a,Legeza-2018a,Shapir-2019,Barcza-2021b,Baiardi-2020}.
In the DMRG method the eigenstate of the Hamiltonian is approximated by a matrix product state (MPS) wavefunction, that is optimized iteratively during the so called DMRG sweeps~\cite{Schollwock-2005, Schollwock-2011}.
Local (1-site or 2-site) optimization steps lead to the iterative diagonalization of the so-called effective quantum many body Hamiltonian ~\cite{Schollwock-2011} that is carried out by the L\'anczos or Davidson algorithms and which corresponds usually to 85-90\% of the total execution time. The effective many body Hamiltonian is built from
renormalized block operators that are formed via the course of the network contraction procedure
which is responsible for another 5-10\% of the total execution time. 
The dimension of the virtual indices of the MPS matrices, usually called as the DMRG bond dimension, $D$, determines the accuracy of the calculations and also the required computational complexity. The overall computational cost of the DMRG for models with generic two-body interactions described by Eq.~(\ref{eq:ham}) scales as $D^3N^4$ where $N$ stands for the number of modes, i.e., for the system size. The memory requirement is proportional to $D^2N^2$.  The details of the algorithm can be found in various review articles \cite{Schollwock-2005,Schollwock-2011,Noack-2005,Szalay-2015a,Chan-2008,Orus-2014,Baiardi-2020}.

When interactions are local, in the sense that interactions are restricted to nearest or next nearest neighbours of the underlying lattice only, the number of nonzero terms in Eq.~(\ref{eq:ham}) is largely reduced.
For example, if one considers the two-dimensional spin-1/2 Hubbard model including only particle hopping, $t$, and on-site Coulomb interaction, $U$, Eq.~\ref{eq:ham} simplifies to
\begin{equation}
\mathcal{H} = \sum_{\langle i,j\rangle\alpha} \left( t\,c^\dagger_{i\alpha}c_{j\alpha} + h.c \right) + 
               \sum_i U n_{i\alpha} n_{j\beta}\,,
\label{eq:ham-hubbard}
\end{equation}
where $i$ and $j$ denote lattice sites on a two-dimensional lattice topology, $\alpha,\beta \in \lbrace \uparrow, \downarrow \rbrace$ denote spin of states, and
$\langle i,j\rangle$ mark nearest neighbours on  the lattice. $c_{i,\sigma}^\dag$ and
$c_{i,\sigma}$ are the fermionic creation and annihilation operators,
satisfying the canonical anti-commutation relations 
$\{c_{i,\sigma},c_{j,\sigma'}\}=0$ and 
$\{c_{i,\sigma}^\dag,c_{j,\sigma'}\}=\delta_{i,j}\delta_{\sigma,\sigma'}$.  
Here $n_{i\alpha}$ is the occupation operator given as $n_{i\alpha}=c^\dagger_{i\alpha}c_{i\alpha}$.
An other simple lattice model is the spinless fermionic model, where
the spin degrees of freedom is also omitted, and whose Hamiltonian reads as
\begin{eqnarray}
\mathcal{H} &=& \sum_{\langle i,j\rangle} \left( t\,c^\dagger_ic_j + h.c \right) + 
              \sum_{\langle i,j\rangle'} \left( t^\prime\,c^\dagger_ic_j + h.c \right)+ \nonumber \\ & &
               \sum_{\langle i,j\rangle} V n_i n_j\,,
\label{eq:ham-spinless}
\end{eqnarray}
where $\langle i,j\rangle$ and $\langle i,j\rangle'$ denote nearest and next nearest neighbours, respectively, on a two-dimensional square lattice.

Boundary conditions play an important role for lattice models, as finite size corrections take different scaling forms~\cite{Cabrera-1987,Nakano-2007,Shibata-2011,Sorella-2015}.
Usually the fastest convergence towards the infinite size limit can be achieved by enforcing periodic boundary condition in both directions. On the other hand, periodic boundary conditions
impose serious drawbacks in the efficiency of the DMRG ~\cite{Noack-1994}, because periodicity introduces long-ranged couplings between sites that are far away from each other in the one-dimensional MPS chain. In addition, when two-dimensional systems are mapped to the one-dimensional topology of the DMRG, local interactions also become non-local. 
Therefore, Hamiltonian given by Eqs.~\ref{eq:ham-hubbard} and \ref{eq:ham-spinless} cannot be considered as a local one in a DMRG calculation. The corresponding MPO bond dimension for models in two and three spatial dimension systematically increases with system size, albeit slower if compared to the general long-ranged model of Eq.~\eqref{eq:ham}. Nevertheless, the algorithmic non-locality of two and three dimensional lattice models makes our approach, that is described below, more and more beneficial in higher dimensional lattices.

\subsection{Joint optimization on the MPS and Grassman manifolds}

The DMRG method provides the wave function in the MPS representation,
\begin{equation}
\ket{\psi} =
\sum_{\substack{\alpha_1,\ldots,\alpha_d\\\in\{0,\downarrow,\uparrow,\downarrow\uparrow\}}} A^{\alpha_1}_{[1]}\cdots A^{\alpha_d}_{[d]}\ket{\alpha_1,\dots,\alpha_d},
\label{eq:DefinitionMPS}
\end{equation}
where the component tensors are $A_{[i]}^{\alpha_i}\in \mathbb{C}^{D_{i-1}\times D_i}$,
with bond dimensions $D_i$, and $D_0 = D_d = 1$.
Although every state vector can be factorized to an MPS form by applying consecutive Schmidt-decompositions~\cite{Schmidt-1907, Vidal-2003b}
using sufficiently large bond dimensions,
however, the sufficient dimensions scale exponentially with the system size in the generic case.
The restriction of the bond dimensions to a fixed value $D$ confines the possible state vectors to a sub manifold of the full state space.
We can then approximate an eigenstate of the Hamiltonian \eqref{eq:ham} within this sub-manifold by the DMRG algorithm, which, being an alternating least square method, optimises the entries of the MPS tensors $A_{[i]}$ iteratively 
\cite{Ostlund-1995,Verstraete-2004a,Verstraete-2004b, Legeza-2014,Szalay-2015a},
leading to a variational treatment of the eigenvalue problem of the Hamiltonian \eqref{eq:ham}.

Utilizing a single particle unitary mode transformation $U \in \mathrm{U}(N)$,
a linear transformation of a set of fermionic annihilation operators $\{c_{i,\sigma}\}$ to a new set $\{d_{i,\sigma}\}$ satisfying the canonical anti-commutation
relations can be obtained, i.e., $c_{i,\sigma} = \sum_{j=1}^d U_{i,j,\sigma} d_{j,\sigma}$. 
We note that in the presented system
it is not necessary to use different unitaries for spin up and down, $U_{i,j,\uparrow}=U_{i,j,\downarrow}$,
however, the implementation is applicable for the unrestricted case too.
Under this transformation, 
the many body representation $G(U)$ can also be expressed on the Fock space \cite{Krumnow-2016},
by which a fermionic wave function
$\ket{\psi(\id)}$ transforms to $\ket{\psi(U)} = G(U)^\dagger\ket{\psi(\id)}$
and the Hamiltonian written in terms of the transformed modes 
by $H(U) = G(U)^\dagger H G(U)$. We note that generic Hamiltonians of form Eq.\eqref{eq:ham} keep their form under such transformation, while simpler Hamiltonians of Eqs.\eqref{eq:ham-hubbard} and \eqref{eq:ham-spinless} lose their simpler structure and are transformed again to the generic form \eqref{eq:ham}.

In the course of our implementation of the DMRG algorithm,
the unitary $U$ is constructed iteratively from two-mode unitary operators
by sweeping through the network. This implementation has the benefit, that the optimal unitaries can be constructed parallel to the DMRG optimization of the eigenstate.
At each micro-iteration step, the half-R\'enyi block entropy
$S_{1/2}(\rho_{\{1,2,\dots,k\}}) = 2\ln(\Tr\sqrt{\rho_{\{1,2,\dots,k\}}})$
is minimized by a two-mode rotation.
(Here $\rho_{\{1,2,\dots,k\}}$ is the density operator of the first $k$ modes \cite{Szalay-2021,Boguslawski-2013}.)
In practice, when turn to numerical simulation including mode optimization,
it is favourable not to transform the operators themselves to keep robustness.
Rather it is practical to perform mode optimization
in terms of the parameters $T_{ij}^{\alpha\beta}$ and $V_{ijkl}^{\alpha\beta\gamma\delta}$ in the Hamiltonian \eqref{eq:ham}. ~\cite{Murg-2010a}

The optimization based on local (two-mode) unitaries often converges to a sub-optimal local minima, therefore we combine it with a global reordering of modes also.~\cite{Krumnow-2016}
At the end of the last DMRG sweep, 
the one-mode entropies\cite{Legeza-2003b} $s_i$, 
the two-mode mutual informations\cite{Rissler-2006}; $I_{i,j}:=s_i+s_j-s_{i,j}$,
the total correlation\cite{Legeza-2004b} $I_\text{tot} = \sum_i s_i$,
the correlation distance $I_\text{dist} = \sum_{i,j} I_{i,j} \vert i - j\vert^2$,
the one-particle reduced density matrix $\gamma_{i,j} = \langle c^\dagger_j c_i \rangle$,
and the occupation number distribution $\langle n_i\rangle$ 
are calculated (where $i,j\in\{1,\ldots,d\}$).
Here, $s_i= -\Tr (\rho_i \ln \rho_i)$ and $s_{i,j}= -\Tr(\rho_{i,j}\ln\rho_{i,j})$
are the von-Neumann entropies of the one- and two-mode reduced density operators $\rho_i$ and $\rho_{i,j}$ \cite{Szalay-2015b}.
The eigenvalues and eigenvectors of the one-particle reduced density matrix
$\gamma$ define the natural occupation numbers $\lambda_i$, and the natural modes (NO).
An optimized ordering of modes along the
tensor network is calculated from the mutual informations $I_{i,j}$, using the Fiedler vector approach \cite{Barcza-2011};
a new {complete active space vector} is calculated from the entropies $s_i$ for the {dynamically extended active space} (DEAS) procedure \cite{Legeza-2003b};
and a new Hartree--Fock (HF) reference configuration is calculated from the occupations $\langle n_i\rangle$. 
These, together with the final rotated interaction matrices, are all used
as inputs for the subsequent mode transformation macro-iteration. To reach sufficient convergence of the optimized basis, several tens of macro iterations are performed, and each macro-iteration contains $5-10$ sweeps of DMRG and two-site unitary micro-iterations with low-to-medium values of bond dimensions $D \lesssim 256$. After an acceptable optimized mode set is found, high precision (large $D$) DMRG calculations are performed without further mode optimization to find eigenstates of the Hamiltonian with high accuracy.
For more details on mode transformation and its applications we guide readers to previous works.~\cite{Krumnow-2016,Krumnow-2019,Krumnow-2021,Mate-2022,Petrov-2023}
To fix notation, we refer by $D_{\rm mo}$ and $D_{\rm opt}$ to the bond dimension used during and after mode optimization procedure, respectively.

\section{Spinless fermion model}
\label{sec:spinless}

In this section we present numerical results together with scaling analysis for the half filled spinless fermion model (Eq.~\eqref{eq:ham-spinless})
on a torus geometry including nearest and next nearest neighbor hopping and nearest neighbor Coulomb interactions. 
This simple model already imposes serious challenges to understand the nature of emerging quantum phases and phase transitions. When nearest neighbor hopping is neglected, i.e., $t^\prime=0$ for half filling, the charge density wave order parameter, $C_{\rm CDW}$, takes on a finite value for arbitrary small Coulomb interaction, as has already been shown in Ref.~\cite{Krumnow-2021}. However, when next-nearest-neighbor hopping is also included, the two-dimensional Fermi surface becomes distorted, and interesting physics emerges.
In general, for small perturbations, the Fermi surface changes and Umklapp scattering dies out,
but the so-called perfect nesting survives, i.e., 
one segment of the Fermi surface can be connected to another segment of the Fermi surface via a reciprocal lattice vector~\cite{Solyom-2011}. For stronger interactions, however, even perfect nesting is destroyed. The nature of the Mott-Hubbard transition in the absence of Umklapp scattering even in one-spatial dimension is a subject of active research and has been explored only recently~\cite{Gebhard-2021}, in the context of the $1/r$-Hubbard model. In this model, there is only a single Fermi point, and the gap opens linearly at a finite Coulomb interaction.
When both local and nearest-neighbor Coulomb interactions are included, a Luttinger-liquid phase emerges in an extended region between the CDW insulator and the Mott-Hubbard insulator phases.~\cite{Gebhard-2023} Analogously, the two
dimensional $t-t^\prime-V$ spinless model is also expected to possess a rich phase diagram.

In the current work, we focus mainly on the technical aspects of our numerical procedure. Thus only the charge density wave order parameter and the quantum information motivated half-chain block entropy will be analyzed for a selected finite value of $t^\prime=0.4$ as a function of $V$.
In addition, for a detailed scaling analysis of the computational complexity, we fix further model parameters to $t=1$, $t^\prime=0.4$ and $V=0.8$, as this point lies in the vicinity of a quantum phase transition, which poses a great challenge to acquire accurate numerical results. A detailed analysis of the two-dimensional phase diagram, together with further analyses of the emerging quantum phases will be part of a subsequent publication. \cite{Noack-2023}

Our software package~\cite{dmrg-budapest} is developed for more than three decades and it has a very flexible structure leading to several highly specialized modules to solve computational tasks in condensed matter theory, quantum chemistry, nuclear physics and in quantum information theory among many others. The main code parts are fully scalable with the number of computational units even on heterogeneous architectures. This is due to new mathematical
models for parallelization, based on self-managed thread scheduling and hierarchical task managements, together with a
new virtual memory management model based on graph theory, which minimizes redundancy between IO operations, substantially increases spatial locality and eliminates the over-reliance on driver level memory
mappings.~\cite{Menczer-2023a,Menczer-2023b,Menczer-2023c}  Computational performance using message phrasing protocol (MPI), openMP, multitherading and GPU are monitored according to various criteria, together with resource usage and communication overhead. These altogether form the basis of our analysis discussed in the following sections.

\subsection{Computational complexity: real space versus optimized basis}
\label{ssec:complexity}

When the full bond dimension is kept, i.e. no truncation is enforced, a unitary transformation on the Hamiltonan leaves the energy spectrum intact. For truncated bond dimension, however, better or worse representations can be obtained via mode optimization in terms of computational complexity. 
Combination of the DMRG with self consistent field (SCF) method based on energy gradient would be a straightforward procedure to obtain an optimized basis, but it requires the calculation of the one- and two-particle reduced density matrices that are unfortunately computationally very expensive \cite{Chan-2008,Murg-2010a}. 
The R\'enyi-entropy based optimization scheme based on entanglement reduction, on the other hand, is a much cheaper method, but in general there is no guarantee that it ultimately leads to energy minimization. Nevertheless, for a large class of problems, also those studied in this work, it provides a powerful tool to attack high dimensional strongly correlated problems~\cite{Krumnow-2016,Krumnow-2019,Krumnow-2021,Mate-2022,Petrov-2023}.

\begin{figure}
    \centering
    \includegraphics[width=0.48\textwidth]
    {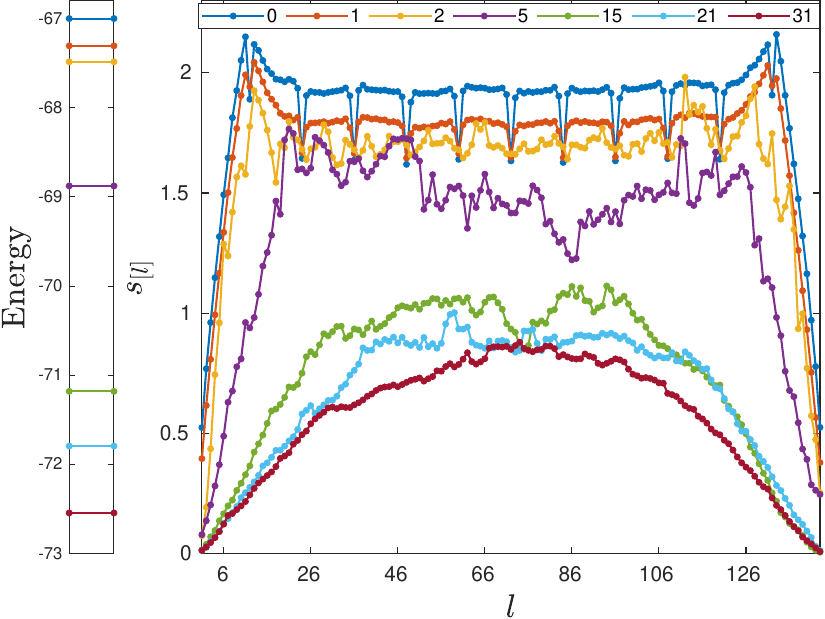}
    \caption{
    Ground state energy (left) and block entropy as a function of the left block size, $\ell$ (right) for some selected mode transformation macro iteration cycles, $0,1,2,5,15,21,31$, with fixed bond dimension $D_{\rm mo}=80$ for the
    half-filled $12\times 12$ two-dimensional spinless fermionic model with model parameters given Fig.~\ref{fig:E_CC_rs_opt}.
    }
    \label{fig:modetrafo}
\end{figure}

\begin{figure}
    \centering
    \includegraphics[width=0.48\textwidth]{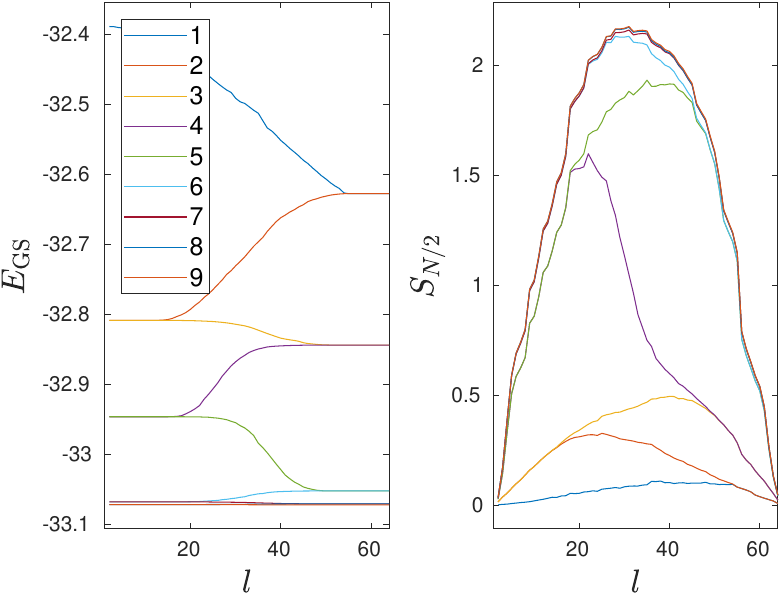}
    \caption{
    Ground state energy (left) and block entropy (right) as a function of left block size, $\ell$, with increasing DMRG sweeps 
    for the half-filled $8\times 8$ two-dimensional spinless fermionic model parameters given in Fig.~\ref{fig:E_CC_rs_opt} using optimized basis, $D_{\rm l}^{(\rm start)}=1024, D_{\rm r}^{(\rm start)}=1024, 
    D_{\rm opt}=4096$ and $\varepsilon_{\rm Davidson}=10^{-7}$.
    }
    \label{fig:energy}
\end{figure}

\begin{figure}
    \centering
    \includegraphics[width=0.48\textwidth]{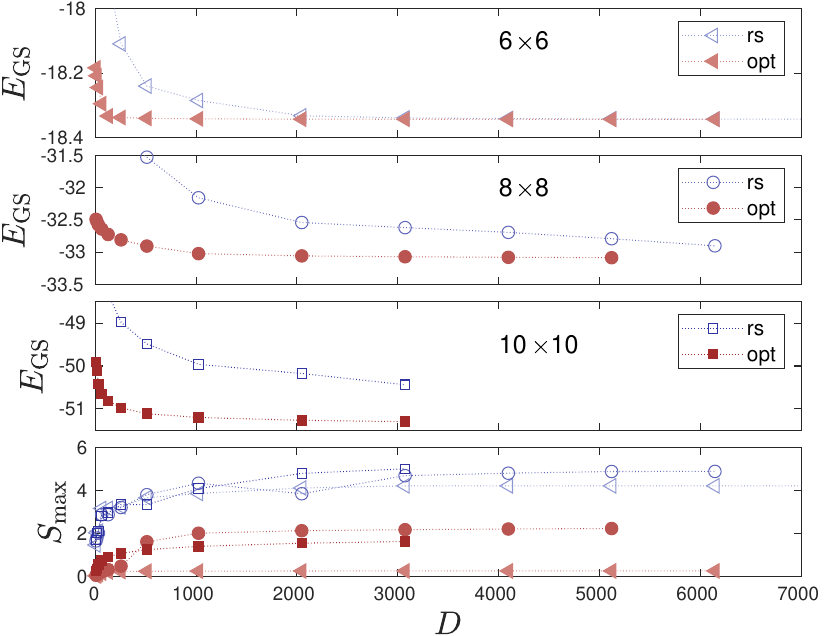}
    \caption{Convergence of the ground state energy and the block entropy as a function of bond dimension using the real space and the optimized basis for
    $6\times 6$, $8\times 8$ and $10\times 10$ lattices and for
    model parameters given in Fig.~\ref{fig:E_CC_rs_opt} and $\varepsilon_{\rm Davidson}=10^{-5}$. 
    }
    \label{fig:Energy_rs_opt}
\end{figure}

In Fig.~\ref{fig:modetrafo} we show result of the R\'enyi entropy based mode optimization procedure for a fixed bond dimension value, $D_{\rm mo}=80$,
for the half-filled $12\times 12$ two-dimensional spinless fermion model with $t=1, t^\prime= 0.4$ and $V=0.8$ on a torus geometry.
The systematic decrease in the ground state energy together with the dramatic drop in the block entropy is obvious. Note that the ground state energy obtained by mode transformation with $D_{\rm mo}=80$ is already far below the one calculated via the real space basis with $D_{\rm rs}=3072$ (see Fig.~\ref{fig:E_CC_rs_opt}).
Therefore, DMRG calculation for a given $D_{\rm opt}$ value using the optimized basis, which basis has been determined beforehand using a bond dimension $D_{\rm mo} \ll D_{\rm rs}$ , is expected to provide significantly more accurate result than with the conventional real space basis.
A typical outcome of such calculation is shown in Fig.~\ref{fig:energy}, i.e., the convergence of the ground state energy and the block entropy as a function of DMRG sweeping.
Although, the block entropy takes much larger values for large $D_{\rm opt}$ compared to the profile optimized with very limited $D_{\rm mo}$ value, i.e. it captures more correlations, the resulting profile, nevertheless, remains almost symmetric. 
The ground state energy and the maximum value of the block entropy is shown in Fig.~\ref{fig:Energy_rs_opt} for the $6\times 6$, $8\times 8$ and $10\times 10$ lattice sizes as a function of $D$. This clearly shows that the optimized modes allow for reaching the same accuracy with significantly lower bond dimension values corresponding to much lower block entropy values. 

Our results are in agreement with previous findings of Ref.~\cite{Krumnow-2019,Krumnow-2021,Mate-2022,Petrov-2023}, but the more general form of the Hamiltionan after mode optimization, Eq.~\ref{eq:ham}, impose serious technical challenges. Although, the block entropy and thus the DMRG/MPS bond dimension is reduced tremendously, the number of terms in the Hamiltonian and, consequently, the bond dimension of its matrix product operator (MPO) representation increases significantly (compare Eqs.~(\ref{eq:ham-hubbard}-\ref{eq:ham-spinless}) with Eq.~(\ref{eq:ham})). Since the overall scaling of the computational complexity of the DMRG for the general form of the Hamiltionian given by Eq.~(\ref{eq:ham}) 
is $D^3N^4$ it remains mandatory to
demonstrate if the reduction in $D$ can overcompensate the increase in the exponent
connected to $N$ when the locality of the Hamiltonian is lost. 
Note that for the two- and higher dimensional systems the locality of the Hamiltonian is already lost when mapped to the one dimensional DMRG topology, thus the increase in the MPO bond dimension is less than for a one-dimensional model.
Here we also remark that
for a given DMRG sweep the scaling is only 
$D^3N^3$ ~\cite{White-1999}, and the extra factor related to system size reflects the number of sweeps required to reach convergence. This latter one is highly problem dependent and the worst scenario holds for models obeying volume law ~\cite{Ehlers-2015}.

\begin{figure}
    \centering
    \includegraphics[width=0.48\textwidth]{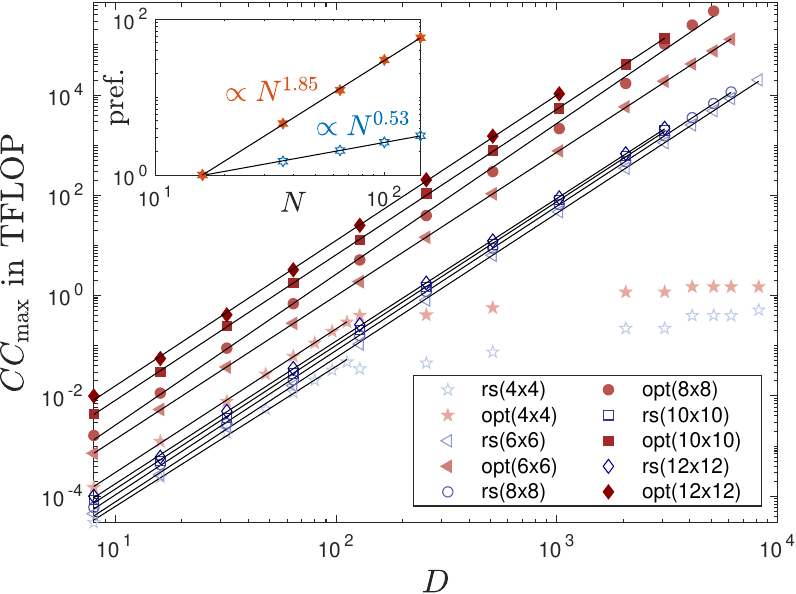}
    \caption{Maximum value of the total computational complexity measured in TFLOP for diagonalization of the effective Hamiltonian as a function of bond dimension using the real space and the optimized basis for model parameters given in Fig.~\ref{fig:E_CC_rs_opt} and $\varepsilon_{\rm Davidson}=10^{-5}$. 
    The solid lines are results of ﬁrst-order polynomial ﬁts leading to exponents 
    $\nu_{\rm rs}(4\times 4)=2.79$, 
    $\nu_{\rm opt}(4\times 4)=2.86$,
    $\nu_{\rm rs}(6\times 6)=2.85$, 
    $\nu_{\rm opt}(6\times 6)=2.85$,
    $\nu_{\rm rs}(8\times 8)=2.9$, 
    $\nu_{\rm opt}(8\times 8)=3.2$,
    $\nu_{\rm rs}(10\times 10)=2.87$,  
    $\nu_{\rm opt}(10\times 10)=2.91$,
    $\nu_{\rm rs}(12\times 12)=2.85$, and 
    $\nu_{\rm opt}(12\times 12)=2.90$.
    The inset shows the scaling of the prefactor as a function of system size $N$ where the fitted exponents lead to 0.53 and 1.85 for the real space and for the optimized basis, respectively.
    }
    \label{fig:CCmax_rs_opt}
\end{figure}

In Fig.~\ref{fig:CCmax_rs_opt} the maximum value of the computational complexity (number of floating point operations), $CC_{\rm max}$, measured via the diagonalization procedure in TFLOP is
summarized for the real space and for the optimized basis as a function of bond dimension on a double logarithmic scale for various systems sizes and for model parameters given in Fig.~\ref{fig:Energy_rs_opt}.
Solid lines are result of first order polynomial fits.
It is clearly seen that data points lie along a straight line. 
Here we remark that for the $4\times4$ model the exact solution is already recovered with $D=2^7=128$ thus a  saturation in $CC_{\rm max}$ becomes apparent for larger $D$ values up to $D=2^{13}$. 
The fitted exponents for the real space basis are in the range of $3\pm 0.2$ confirming the expected $D^3$ scaling. 
Data points for the optimized basis
also lie along a straight lines corresponding to fitted exponents in the range of $3\pm 0.2$. Thus the overall scaling with $D$ does not change due to mode optimization, as expected, but
lines shift toward larger $CC$ values due to increase in the MPO bond dimension.
Switching from real space to mode optimized basis the scaling of such prefactor is expected to change from $\sqrt N$ to $N^2$.
In fact, as shown in the inset of Fig.~\ref{fig:CCmax_rs_opt}, the fitted exponents of the prefactor, as a function of system size $N$, lead to $0.53$ and $1.85$ for the real space and for the optimized basis, respectively. 
Therefore, naively, one could render the optimized basis behind real space basis, but after recalling that the same accuracy in energy can be reached with significantly lower $D$ values, this is no longer true.
In fact, as can be seen in Fig~\ref{fig:E_CC_rs_opt} the same accuracy in energy can be reached with $D_{\rm opt}=512$ as with $D_{\rm rs}\simeq 3000$ for $6\times 6$ and with
$D_{\rm opt}=512$ as with
$D_{\rm rs}\simeq 5500$ for $8\times 8$ model, respectively \footnote{Here by $D_{\rm opt}$ we refer to the bond dimension used after the mode optimization procedure. Mode optimization were performed at fixed bond dimension $D_{\rm mo} = 80$.} The saving in computational complexity increases significantly with system size,
for example, for $10\times 10$ and $12\times 12$ this gets close to four orders of magnitude. 

We conclude, that the computational complexity for a given accuracy can be reduced by several orders of magnitude, depending on system size, via mode optimization.
Since mode optimization is performed with low bond dimension, the computational complexity of the optimization procedure is negligible compared to the saving when the post-modetransformation DMRG calculations are performed using the optimized basis. The data for $10 \times 10$ and $12 \times 12$ show that mode optimization can allow for accuracy that is far beyond the scope of real space DMRG.

While in Fig. \ref{fig:modetrafo}, for simplicity, a fixed bond-dimension $D_{\rm mo} = 80$ has been used for the mode optimization procedure, in practice, usually we start mode optimization with an even lower bond dimension value, performing a finite number of mode transformation macro iteration cycles each comprising DMRG calculations with finite number of sweeps, and this procedure is repeated by systematically increasing the bond dimension\cite{Krumnow-2021,Petrov-2023}. This approach further reduces the calculation time of the optimization procedure, and also speeds up convergence at initial sweeps by avoiding sticking to local minima of the block entropies.
A typical setting we used is $D_{\rm mo}=[8,16,64,80]$, ${\rm Sweep}_{\rm mo}=[11,11,9,7]$ and ${\rm Iter}_{\rm mo} = [9,9,7,5]$.

\subsection{Performance: real space versus optimized basis} 
\label{ssec:performnace}

Our arguments and conclusion presented in Sec.~\ref{ssec:complexity} can be further improved if we also consider the underlying power in state of the art hardware and software technologies, i.e., massive parallelization via message passing interface (MPI) and graphical processing units (GPUs). 
Here we demonstrate that our hybrid CPU-multiGPU solution ~\cite{Menczer-2023a,Menczer-2023b} has the potential to reduce significantly the effective exponent connected to $D$, thus reducing further the overall wall time of the calculations drastically. 

\begin{figure}
    \centering
    \includegraphics[width=0.48\textwidth]{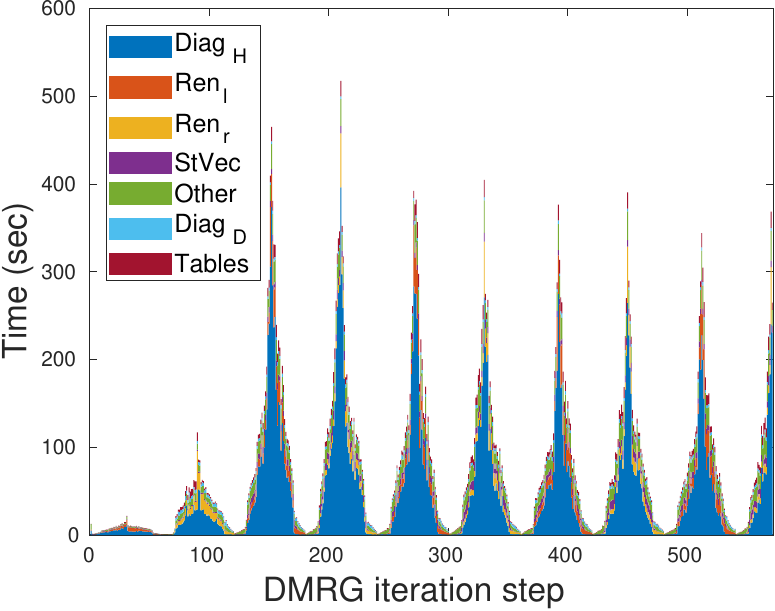}
    \caption{
    Contribution of the most expensive functions to the total time measured in seconds as a function of the DMRG iteration step for model and DMRG parameters given in Fig.~\ref{fig:E_CC_rs_opt}.
    The calculation has been performed on a dual
    AMD EPYC 7702 CPUs with $2\times 64$ cores compiled with eight NVIDIA A100-SXM4-40GB devices.  
    }
    \label{fig:perf_all}
\end{figure}

First we present detailed performance analysis for a single DMRG calculation 
for the $8\times 8$ lattice, for model parameters given in Fig.~\ref{fig:E_CC_rs_opt}, using the R\'enyi entropy optimized basis, $D_{\rm l}^{(\rm start)}=1024, D_{\rm r}^{(\rm start)}=1024, D=4096$ and $\varepsilon_{\rm Davidson}=10^{-7}$.
From technical point of view, for the spinless fermion model only the total particle number can be used as a conserved quantum number, thus the underlying DMRG matrix and tensor algebra decomposes into fewer but larger sectors compared to spinful models.
~\cite{Menczer-2023a,Menczer-2023b} 
In Fig.~\ref{fig:perf_all} we show the contribution of the most expensive functions to the total time measured in seconds as a function of the DMRG iteration step.
Here Diag$_H$ denotes the time spent on the diagonalization of the effective Hamiltonian, Ren$_{\rm l}$ and Ren$_{\rm r}$ the renormalization step for the left and right block via the forward and backward sweep, respectively. The overhead to prepare the quantum number decomposed based matrix and tensor algebra is labeled by $\rm{Tables}$, the diagonalization of the reduced density matrix, i.e., SVD step by Diag$_D$, the wave function transformation to generate starting vector for the diagonalization step by $\rm{StVec}$, and all other utility components by $\rm{Other}$.
The calculation has been performed on a dual AMD EPYC 7702 CPUs with 2 × 64 cores compiled with eight NVIDIA A100-SXM4-40GB devices.  
It is clear that even after full GPU parallelization 80-90\% of the total time in a given DMRG iteration step is allocated for the diagonalization of the effective Hamiltonian. We also note that a factor 25-30 speedup is already gained with respect 
to a fully parallelized CPU only implementation~\cite{Menczer-2023a}.
Therefore, in the following we restrict our analysis on the diagonalization procedure. Here we remark, that via the sweeping procedure the renormalized operators, contracted network components, must be stored and reused in subsequent iteration steps, which IO time can be significant if disks are used for that purpose.

\begin{figure}
    \centering
    \includegraphics[width=0.48\textwidth]{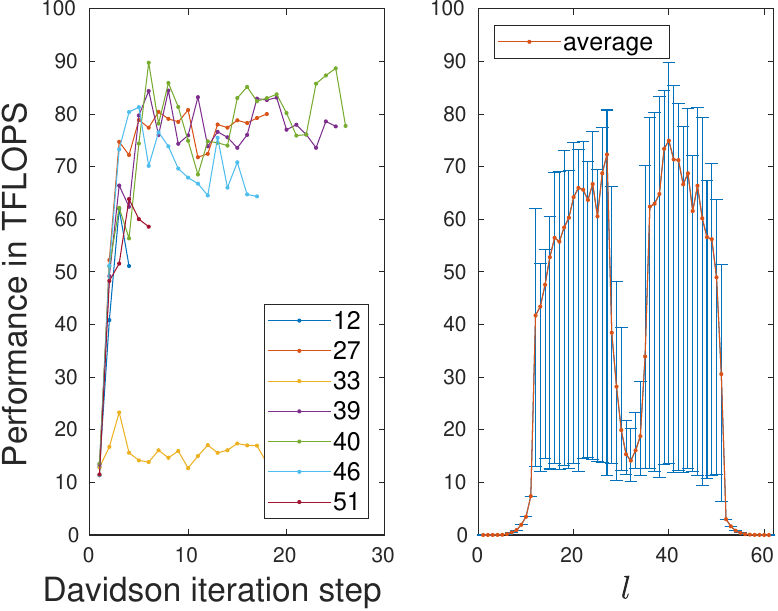}
    \caption{(a) Performance in TFLOPS measured via the eight GPU accelerated Davidson diagonalization procedure for some selected DMRG iterations steps of the fifth DMRG sweep for model and DMRG parameters given in Fig.~\ref{fig:E_CC_rs_opt}. Numbers in the legend stand for size of the left block size, $l$.
    (b) Minimum, maximum and average performance measured in TFLOPS for the fifth DMRG sweep as a function of left block size, $l$, for model and DMRG parameters given in Fig.~\ref{fig:E_CC_rs_opt}.
    }
    \label{fig:perf_average}
\end{figure}

To provide more insights, in Fig.~\ref{fig:perf_average}(a) we present performance measured
in TFLOPS via the Davidson diagonalization procedure for some selected DMRG iteration steps of the fifth DMRG sweep for model and DMRG parameters given in Fig.~\ref{fig:E_CC_rs_opt}. Numbers in the legend stand for size of the left block size, $l$.   
It is clearly seen that the number of matrix-vector operations fluctuates heavily depending on the DMRG iteration steps, while the performance for a given DMRG iteration step reaches a high values with increasing Davidson steps. Here we remark, that for the first few Davidson iteration steps, the IO overhead to transfer data form host to device (H2D IO) and the overhead for our dynamic scheduler to be optimized~\cite{Menczer-2023b} reduces the performance to lower initial values. 
The corresponding minimum, maximum and average performance values are collected and presented in Fig~\ref{fig:perf_average}(b). Note that the FP64 limit of NVIDIA~\cite{nvidia}, 76 TFLOPS, is reached by most of the iteration steps and the average performance is effected only marginally by the low initial performance values of the Davidson procedure.

To complete our analysis, in Fig.~\ref{fig:Perf_rs_opt} we compare the largest performance and wall time as a function of DMRG bond dimension for calculations obtained with real space and with optimized basis. 
For real space DMRG, the MPO bond dimension is much lower, therefore, the efficient matrix and tensor algebra with SBATCH \cite{Menczer-2023b} cannot be utilized as efficiently as for the optimized basis.
This leads to a relatively poor performance 
as even for the eight GPU accelerated diagonalization procedure the largest performance is about 10-14 TFLOPS (see Fig.~\ref{fig:Perf_rs_opt} left panel). In contrast to this, for the optimized basis the underlying GPU power can be utilized very efficiently. The sharp increase in performance up the FP64 ceiling is obvious while a saturation for larger $D$ values becomes apparent. Nevertheless, our results already demonstrate the utilization of the highly specialized NVIDIA tensor core units (TCUs)~\cite{Menczer-2023b}. For larger system sizes the sudden increase in performance gets even more pronounced as has also been demonstrated for quantum chemical calculations~\cite{Menczer-2023b}.

The significant difference in reachable performance for the real space and for the optimized basis also determines the total wall time.
The total time spent on the eight GPU accelerated diagonalization of the effective Hamiltonian including host to device, H2D, and device to host, D2H, IO measured for nine DMRG sweeps is presented in the right panel of Fig.~\ref{fig:Perf_rs_opt} on a double logarithmic scale as a function of bond dimension for the $n \times n$ half-filled spinless model with $n=6,8,10,12$ with model and DMRG parameters given in Fig.~\ref{fig:E_CC_rs_opt}.
It is apparent in the figure that via massive parallelization the total diagonalization time for nine sweeps for a given $D$ value and system size is almost the same for the real space and for the optimized basis.
For the $6\times 6$ lattice size the calculations are even faster for the optimized basis, for $8\times8$ almost the same while for large $N$ calculations with the same D gets more expensive, thus the difference increases in system size. 
Recalling, however, that for $6\times 6$ lattice the same accuracy is obtained with $D=512$ for the optimized basis taking two minutes as with $D=3000$ for the real space basis taking 20 minutes already one order of magnitude reduction can be achieved in wall time for this small system size.
Similarly for $8\times 8$, the corresponding numbers are $D=512$ and seven minutes for the optimized basis while the matching calculation in energy with $D\simeq 5500$ takes about 200 minutes, i.e., the reduction of wall time is almost two orders of magnitude. For larger system sizes such comparison is not possible as the real space approach has been unable to reach the same accuracy as the optimized basis has reached already for the lowest considered bond dimension $D_{\rm opt} = 64$. Nevertheless, our data indicate that the speedup is over two orders of magnitude in these cases. 
Note that data are measured for the eight GPU accelerated implementation for all system sizes.

Here we also remark that the saturation in performance will be lifted and even eliminated via our new kernel utilizing also NVIDIA fast device to device (D2D) NVLINK and message phrase interface (MPI)~\cite{Menczer-2023c}. Therefore, further reduction in wall time for the optimized basis can be achieved for a much broader range of the bond dimension.

\subsection{Quantum phase transition at finite Coulomb interaction} 
\label{ssec:qpt_coulomb}
\begin{figure}
    \centering
    \includegraphics[width=0.48\textwidth]{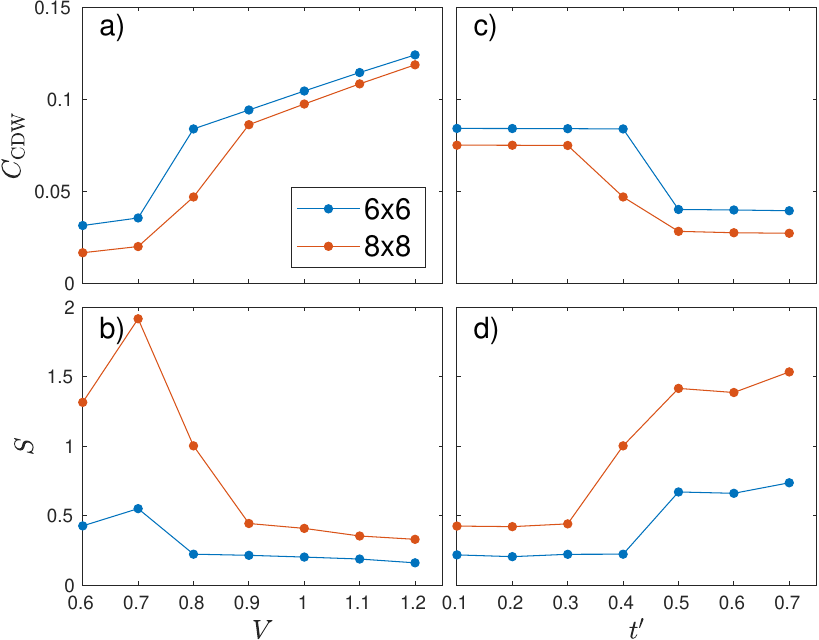}
\caption{(a) Charge density wave order parameter, $C_{\rm CDW}$, 
and (b) half-chain block entropy obtained via the optimized basis
for the $6\times 6$ and for the $8\times 8$ spinless fermion model on a torus geometry as a function of $V$ for finite $t^\prime=0.4$. (c) and (d) the same but as a function of $t^\prime$ for $V=0.8$.}
\label{fig:cdw}
\end{figure}

In addition to groundstate energies, our procedure also allows us to calculate expectation values of operators with much higher accuracy than can be done in the real-space basis. As discussed in Ref.~\cite{Krumnow-2021}, correlation functions, and one-  and two-body reduced density matrices can be obtained very accurately in the optimized basis and can then be transformed back to the original, real-space basis via a sequence of rotations using unitary matrices calculated and stored during mode optimization. Here we remark that the calculation of the computationally very demanding 4-index tensor of the two-body reduced density matrix, $\Gamma_{ijkl}=\langle c^\dagger_i c^\dagger_j c_k c_l\rangle$, can also be accelerated tremendously via our hybrid CPU-multiGPU kernels. By taking a linear combination of proper matrix elements of the back-rotated reduced density matrices, the CDW order parameter, $C_{\rm CDW} =\frac{1}{N^2} \left\langle \left(n_{\rm{even}} - n_{\rm{odd}}\right)^2 \right\rangle$, can, for example, be calculated, where $n_{\rm{even/odd}}$ stands for the total particle number in the even/odd sublattices.
This quantity is shown in Fig.~\ref{fig:cdw}, together with the half-chain block entropy for the optimized basis as a function of $V$ for finite $t^\prime=0.4$ and as a function of $t^\prime$ for $V=0.8$ 
for the $6\times 6$ and $8\times 8$ spinless fermion model on torus geometry. A sudden change in $C_{\rm CDW}$
is apparent at a finite value of $V$ that shifts slightly from $V_{\rm c}=0.7$ to $V_{\rm c}=0.8$ with increasing system size.
The large jump in the block entropy also indicates the anomalous behavior at finite $V_{\rm c}$. Similar behavior is observed as a function of $t^\prime$ at $t^\prime\simeq0.4$ for $V=0.8$.
Further numerical and analytic analysis of the phase diagram is beyond the scope of the present work due to its very 
complex structure and will be part of subsequent work~\cite{Noack-2023}.

\section{Hubbard model}
\label{sec:hubbard}

In this section, we present scaling analysis and benchmark results obtained via the mode optimized non-Abelian $SU(2)$ spin adapted hybrid CPU-multiGPU DMRG method for the half-filled two-dimensional Hubbard model, (\eqref{eq:ham-hubbard}), on a torus geometry. 
Besides the total charge, the spin quantum number can also be kept which leads to significant increase in the number of independent tasks and to smaller sector sizes. In addition, the $SU(2)$ spin symmetry can also be utilized leading to tremendous increase in effective performance without additional computational overhead~\cite{Menczer-2023b}. In this latter case, the DMRG matrix and tensor algebra is reformulated according to non-Abelian quantum numbers, thus the bond dimension $D$ stands for 
the number of renormalized multiplets.
When the corresponding $U(1)$ bond dimension is also indicated it is denoted explicitly as $D_{U(1)}^{\rm eff}$.

Here, we perform mode optimization using $U(1)$ symmetries only, by either enforcing the same unitary for both spin components, thus preserving the $SU(2)$ symmetry of the Hamiltionian, or by optimizing two different unitary matrices for the two spin components independently. We remark, that the $SU(2)$ spin adapted version of the mode optimization procedure is part of our current developments. 
The post mode transformation DMRG calculations with large $D$ values can be performed using $U(1)$ and $SU(2)$ symmetries.

\begin{figure}
    \centering
    \includegraphics[width=0.48\textwidth]{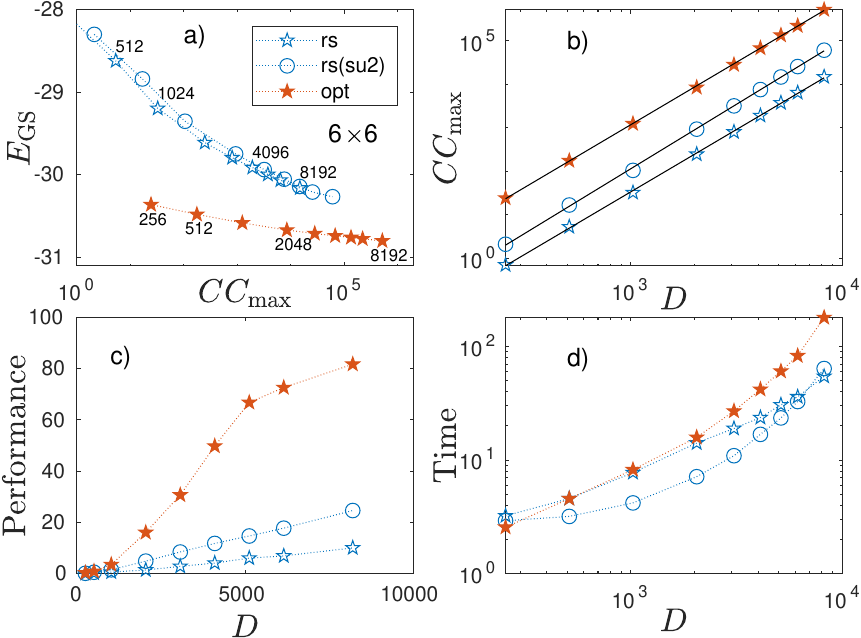}
    \caption{Ground state energy as a function of computational complexity (a), computational complexity in TFLOP (b), largest performance in TFLOPS (c) and total wall time of 9 sweeps in minutes (d) as a function of bond dimension obtained via the real space basis with $U(1)$ and $SU(2)$ spin symmetry and by optimized modes with $U(1)$ symmetry for the half-filled $6\times6$ Hubbard model at $U=4$ on a torus geometry. The fitted exponents in
    (b) are $\nu_{\rm rs}=2.96$ and $\nu_{\rm opt}=2.87$.
    For the $SU(2)$ spin adapted calculations $D$ stands for the number of renormalized multiplets.
    }
    \label{fig:hub_6x6_u4}
\end{figure}
In Fig.~\ref{fig:hub_6x6_u4} we summarize our results for the $6\times 6$ half-filled Hubbard model at $U=4$ using real space basis with $U(1)$ and $SU(2)$ spin symmetries and via mode optimized basis using $U(1)$ symmetries only. The mode optimization has been performed iteratively up to $D_{mo}=128$.
The $U=4$ interaction strength was selected as there are several benchmark results available for torus geometry in the literature obtained by various methods~\cite{Qin-2016,Dobrautz-2019,Wu-2019}.
First we remark that although the $SU(2)$ spin adapted version for the real space basis provides lower energy than the $U(1)$ implementation for a given bond dimension 
(Fig.~\ref{fig:hub_6x6_u4}(a))
it corresponds to enlarged computational complexity (Fig.~\ref{fig:hub_6x6_u4}(b). This is due to larger number of, and more dense, sectors. Nevertheless, it can be parallelized more efficiently resulting in larger performance values as a function of $D$ (Fig.~\ref{fig:hub_6x6_u4}(c)). Therefore, the overall wall time can be scaled below the $U(1)$ implementation for a broad range of the bond dimension (Fig.~\ref{fig:hub_6x6_u4}(d)).  

\begin{figure}
    \centering
    \includegraphics[width=0.48\textwidth]{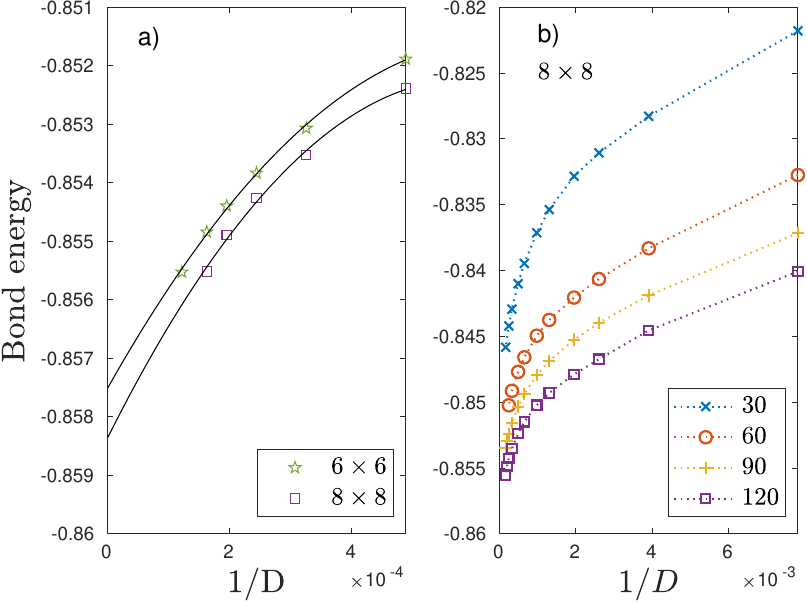}
    \caption{(left) Bond energy as a function of inverse bond dimension for the optimized modes with $U(1)$ symmetry for the half-filled $6\times6$ and $8\times8$ Hubbard models at $U=4$ on a torus geometry. The solid lines are  second order polynomial fits.
    (right) Bond energy as a function of inverse bond dimension using optimized basis via 30, 60, 90 and 120 macro mode optimization cycles 
    for the $8\times8$ system size.}
    \label{fig:hub_bondenergy}
\end{figure}

When mode optimization is performed we have found that a significantly larger reduction in block entropy, together with a significantly lower energy can be achieved when the unitary matrices are optimized independently for the two spin components. Although the $SU(2)$ symmetry is broken for such unrestricted optimization procedure, the low ground state energy could not be recovered via the post mode transformation DMRG calculations even using $SU(2)$ symmetry and large bond dimension when the same angles were enforced during the optimization procedure. We note that with $D_{\rm opt}=256$ DMRG provides lower ground state energy for the optimized basis than for the real space basis with $D=8192$ together with $SU(2)$ spin symmetry, i.e., for an effective $U(1)$ bond dimension around 25000 (Fig.~\ref{fig:hub_6x6_u4}(a)). In addition, a very reliably scaling
with inverse bond dimension can be achieved for the optimized modes (see Fig~\ref{fig:hub_bondenergy}) leading to an extrapolated bond energy $E_{\rm GS}/N=-0.8575(2)$. 
In Table~\ref{tab:hub_6x6_u4}. we summarize our results for the bond energy together with reference data from the literature obtained by various methods. It is evident that our method provides bond energy in the range of the reference data sets and using optimized modes a very reliable $1/D$ extrapolation can be achieved. 
Our result is, in fact, in close agreement with bond energy obtained by AFQMC~\cite{Qin-2016}.
We also remark that unlike DMRG, most of the reference methods are non variational. 
\begin{table}[htb]
\begin{tabular}{|l|c|c|}
\hline
Bond energy &  method  & $\max(D_{U(1)}^{\rm eff}$)\\
\hline
\hline
-0.85625(30) &  iST-FCIQMC\cite{Dobrautz-2019} & -\\
-0.85736(25)  &  AFQMC\cite{Qin-2016} & -\\
-0.79703      &  UHF~\cite{Wu-2019} & -\\
-0.86792      &  p-DMET~\cite{Wu-2019} & -\\
-0.86856      &  DMET~\cite{Wu-2019} & - \\
-0.83894     & DMRG[6140,SU(2)]  & 19748 \\
-0.84067     & DMRG[8192,SU(2)]  & 25088 \\
-0.84333     & DMRG[128,256,U(1)] & 256\\
-0.85553     & DMRG[128,8192,U(1)] & 8192\\
-0.8575(2)   & DMRG extrapolated & -\\
\hline
\end{tabular}
\caption{Bond energy for the half-filled $6\times 6$
Hubbard model for $U=4$ on a torus geometry obtained by various methods. Our notation reads as DMRG[$D_{\rm rs}$, Symmetry] and
DMRG[$D_{\rm mo}$, $D_{\rm opt}$, Symmetry], respectively.
The maximum effective bond dimension is also included in the last column.
Note that iST-FCIQMC, AFQMC, p-DMET, DMET are non variational methods.
}
\label{tab:hub_6x6_u4}
\end{table}

We close our analysis by providing benchmark data for the $8\times8$ half-filled Hubbard model (see Fig.~\ref{fig:hub_bondenergy}) together with extrapolation leading to $E_{\rm GS}/N=-0.8584(2)$.
This value is close, but off by $10^{-3}$ compared to the bond energy obtained by AFQMC~\cite{Qin-2016}. Note, however, that by increasing $D_{\rm opt}$ further one could get even closer to the error free solution and obtain better extrapolation, it would simply require more computational resources. On the other hand, the rate of convergence depends significantly on the quality of the optimized basis, i.e., there is an initial error obtained by mode optimization with fixed rank $D_{\rm mo}$ if not a fully converged data set is used. This effects the $1/D_{\rm opt}$ extrapolation as well, i.e., lower energies with the same $D_{\rm opt}$ can be reached if a better basis is used. This is demonstrated in Fig.~\ref{fig:hub_bondenergy}(b) for the $8\times 8$ system size using optimized modes obtained via 30, 60, 90 and 120 macro mode optimization cycles. 

To gain more insights into how mode optimization is influenced by the nature of the quantum many body wave function, we have repeated our analysis for attractive interactions. By taking a negative value, $U=-4$ , we find that the two angles during mode optimization for the two modes take on very close values. This is not a surprise, as for
attractive interaction, particles with opposite spins like to form pairs, which is in favor of preserving $SU(2)$ symmetry. 
Accordingly, the expectation value of the occupation number operator of both spin-up and spin-down components oscillates with the same profile along the lattice sites. 
In contrast, the ground state of the repulsive ($U>0$) model is characterized by antiferromagnetic correlations, in which electrons of differing spin are in different sublattices (as also confirmed by our numerics through the $\langle n_{i \uparrow/\downarrow} \rangle$ expectation values). Enforcing a spin-independent spatial form of fermionic modes during the mode optimization is not optimal for such states, leading to reduced performance for the repulsive case. We have found, however, that there is very stable and fast convergence in the attractive case with enforced $SU(2)$ symmetry and obtain much lower ground state energies for a given bond dimension. Therefore, we conclude that the optimal unitary transformations that correspond to independent modes depend on the structure of the ground state. Nevertheless, our procedure works in general.

Further algorithmic developments to boost the convergence of the mode optimization protocol
 and to lower such initial error together with additional analysis including other fillings and topologies is under investigation and will be part of our subsequent work. 
 Nevertheless, our procedure is variational, free of sign problem, thus can be applied at general fillings and for frustrated models too.  
Finally, an important advantage of the mode optimized basis is that it can be parallelized very efficiently reaching again 90 TFLOPS in our benchmark calculations (Fig.~\ref{fig:hub_6x6_u4}(c)) reducing wall time to the same range as the real space calculation with $U(1)$ symmetry
(Fig.~\ref{fig:hub_6x6_u4}(d)). In addition, improving performance further via our CPU-multiGPU kernel together with MPI protocol and fast NVIDIA D2D NVLINK will reduce even more the total execution time for larger system sizes and $D$ values. 

\section{Conclusion}
\label{sec:conclusion}

In this work, we have presented a hybrid numerical approach to simulate quantum many body problems on two spatial dimensional quantum lattice models via the non-Abelian ab initio  version of the density matrix renormalization group method on state-of-the-art high performance computing infrastructures. 
We demonstrated for the two dimensional spinless fermion model and for the Hubbard model on torus geometry that altogether several orders of magnitude in computational time can be saved for a given accuracy by performing calculations on an optimized basis and by utilizing hybrid CPU-multiGPU parallelization. At least an order of magnitude reduction in computational complexity, measured directly in FLOP, results from mode optimization while an order of magnitude reduction in wall time is achieved by massive parallelization. 
A detailed scaling analysis of the obtained performance measured in FLOPS as a function of matrix ranks and as a function of system size up to $12\times 12$ lattice topology 
for the spinless model
revealed that more efficient parallelization can be gained for the optimized modes. As a result, even though the computational complexity increases via mode optimization, the 
overall wall time can be reduced to similar magnitude as for the real space basis for a broad range of bond dimension values.

For the Hubbard model we also analyzed the scaling of the various quantities for the real space basis using $U(1)$ symmetry only and for the $SU(2)$ spin adapted version. For the mode optimization we have found that a much better basis can be obtained by optimizing unitary matrices for two spin components independently. After such unrestricted optimization protocol we managed to provide new DMRG lower bond energy values and reliable extrapolations for the non-truncated limit with respect to previous attempts which are, in fact, in good agreement with AFQMC reference data set.

Our numerical results, however, show that mode optimization is a more delicate issue for the Hubbard model and it requires further developments to boost convergence. Nevertheless, mode optimization is crucial to obtain a quasi optimal basis, i.e., to reduce initial error for an optimization on a fixed rank manifold. 
It is worth to note an inbalance in current developments in information technology, namely, the number of computational units increases in a much higher rate than the available size of memory on HPC infrastructures.
Therefore, it is more important to reduce the rank of the matrices and tensors, than minimizing the computational complexity. This latter one can be handled more easily via massive parallization. 

Finally, our approach is variational, free of sign problem and can be applied to more general fillings and topologies and for frustrated systems as well.
It can also be further improved by utilizing message phrase passing (MPI) protocols and NVIDIA fast device to device NVLINK communication. These latter aspects are part of our current developments.

\section*{Acknowledgments}
The authors acknowledge useful discussions with Tam\'as Kozsik, Reinhard Noack, Florian Gebhard, and Jen\H{o} S\'olyom. This work has been
supported by the Hungarian National Research, Development and Innovation Office (NKFIH) through Grant Nos.~K134983 and TKP2021-NVA-04,
by the Quantum Information National Laboratory
of Hungary. \"O.L. acknowledges financial support
by the Hans Fischer Senior Fellowship programme funded by the Technical University
of Munich – Institute for Advanced Study and by
the Center for Scalable and Predictive methods
for Excitation and Correlated phenomena (SPEC),
funded as part of the Computational Chemical Sciences Program FWP 70942 by the U.S. Department of Energy
(DOE), Office of Science, Office of Basic Energy Sciences, Division of Chemical Sciences, Geosciences, and Biosciences at Pacific Northwest National Laboratory.
M.A.W. has also been supported by the Janos
Bolyai Research Scholarship of the Hungarian Academy of Sciences.
The simulations were performed on the national supercomputer HPE Apollo Hawk at the High Performance Computing Center Stuttgart (HLRS) under the grant number MPTNS/44246.

\end{document}